\documentclass[pra,amsmath,twocolumn,showpacs,floatfix]{revtex4}
\pdfoutput=1
\usepackage{calc}
\usepackage{graphicx}
\usepackage{amsfonts} 
\usepackage{esint}
\usepackage{xspace}
\usepackage[utf8]{inputenc}

\usepackage{hyperref}
\hypersetup{
   colorlinks=true,
   linkcolor=black,
   citecolor=black,
   urlcolor=blue,
   pdfauthor={Janus H. Wesenberg},
   pdftitle={Electrostatics of surface-electrode ion traps}
}

\graphicspath{{figures/}{figures/converted/}}

\renewcommand{\vec}[1]{\ensuremath{\boldsymbol{#1}}}
\newcommand{\vhat}[1]{\ensuremath{\hat{\vec{#1}}}}

\newcommand{\topp}[1]{^{({#1})}}

\newcommand{\grad}{\vec{\nabla}}

\DeclareMathOperator*{\re}{Re}
\DeclareMathOperator*{\im}{Im}
\DeclareMathOperator{\bigo}{\mathcal{O}}

\newcommand{\vecc}[1]{\tilde{{#1}}}

\newcommand{\fieldC}{\ensuremath{\mathbb{C}}}
\newcommand{\fieldR}{\ensuremath{\mathbb{R}}}

\newcommand{\ueff}{U_{\text{eff}}}
\newcommand{\upp}{U_p}

\newcommand{\pbsaddle}{\ensuremath{p_s}}
\newcommand{\psaddle}[1]{\ensuremath{p_s\topp{#1}}}
\newcommand{\pgsaddle}[1]{\ensuremath{\bar{p}_s\topp{{#1}}}}
\newcommand{\pasaddle}[2]{\ensuremath{p_s\topp{{#1},{#2}}}}

\newcommand{\usaddle}[1]{\ensuremath{u_s\topp{{#1}}}}
\newcommand{\ugsaddle}[1]{\ensuremath{\bar{u}_s\topp{{#1}}}}
\newcommand{\uasaddle}[2]{\ensuremath{u_s\topp{{#1},{#2}}}}

\newcommand{\dsaddle}[1]{\ensuremath{U_s\topp{{#1}}}}
\newcommand{\dgsaddle}[1]{\ensuremath{\bar{U}_s\topp{{#1}}}}

\newcommand{\vdc}{\ensuremath{v_{\text{c}}}}

\newcommand{\vsg}[1]{\ensuremath{\bar{V}_{#1}}}
\newcommand{\vbg}[1]{\ensuremath{\Theta_{#1}}}
\newcommand{\vag}[1]{\ensuremath{V_{#1}}}

\newcommand{\vsrf}{\vsg{\text{rf}}}
\newcommand{\vsci}[1]{\vsg{\text{c},#1}}
\newcommand{\vscrf}{\vsg{\text{c},\text{rf}}}

\newcommand{\varf}{\vag{\text{rf}}}
\newcommand{\vaci}[1]{\vag{\text{c},#1}}
\newcommand{\vacrf}{\vag{\text{c},\text{rf}}}

\newcommand{\vbrf}{\vbg{\text{rf}}}
\newcommand{\vbci}[1]{\vbg{\text{c},#1}}

\begin{document}
\title{%
  Electrostatics of surface-electrode ion traps
}
%\subtitle{}
\date{\today}

\author{J.~H.~Wesenberg}
\email[]{janus.wesenberg@materials.ox.ac.uk}
\affiliation{Department of Materials, University of Oxford, United Kingdom}
\date{\today}

\begin{abstract}
  Surface-electrode (SE) rf traps are a promising approach to
  manufacturing complex ion-trap networks suitable for large-scale
  quantum information processing. In this paper we present 
  analytical methods for modeling SE traps in the gapless plane
  approximation, and apply these methods to two particular classes of
  SE traps.
  For the SE ring trap we derive analytical expressions for the trap
  geometry and strength, and also calculate the depth in the absence
  of control fields.
  For translationally symmetric multipole configurations (analogs of
  the linear Paul trap), we derive analytical expressions for
  electrode geometry and strength. Further, we provide arbitrarily
  good approximations of the trap depth in the absence of static
  fields and identify the requirements for obtaining maximal depth. 
  Lastly, we show that the depth of SE multipoles can be greatly
  influenced by control fields.
\end{abstract}

%37.10.Gh 	Atom traps and guides
%41.20.Cv 	Electrostatics; Poisson and Laplace equations, boundary-value problems

\maketitle
\begin{figure}
  \centering
  \includegraphics{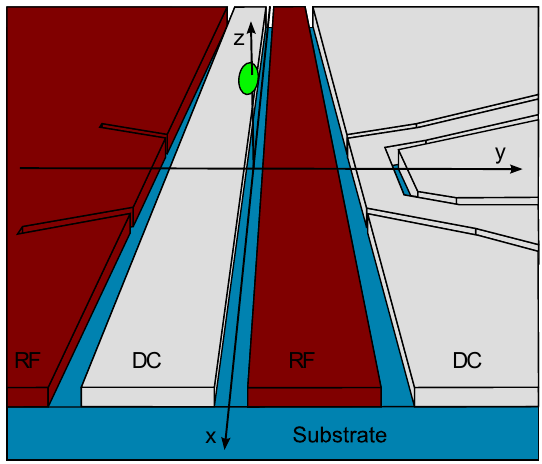}
  \caption{(Color online)
    Cut near the trap center (oval) of a surface-electrode (SE)
    trap, showing the rf and control (DC) electrodes on an insulating
    substrate. Design, including the aspect ratio of the gaps between
    electrodes, corresponds roughly to that of the trap described in
    Ref.~\cite{seidelin06:microf_surfac_elect_trap_scalab}, for an ion
    surface distance of $40\,\text{$\mu$m}$.
    The electrodes extend beyond the illustrated area in all
    directions.  }
  \label{fig:signe_elec}
\end{figure}

Several approaches to large-scale quantum information processing (QIP)
with trapped ions using networks of interconnected rf traps have been
proposed
\cite{kielpinski02:archit_a_large_scale_trap,leibfried07:transport,ospelkaus08:trapped-ion}.
Surface-electrode (SE) ion traps, where the trap electrodes are
located in a single plane as illustrated in Fig.~\ref{fig:signe_elec},
offer a promising approach to constructing such trap networks
\cite{chiaverini05:surfac_elect_archit_for_ion,seidelin06:microf_surfac_elect_trap_scalab,labaziewicz08:suppression}.
SE traps are well suited for microfabrication, and motional heating
rates compatible with QIP have been demonstrated
\cite{seidelin06:microf_surfac_elect_trap_scalab,labaziewicz08:suppression},
an important fact since motional heating is currently a limiting
factor for ion-trap miniaturization
\cite{turchette00:heatin_trapp_ions_from_quant,epstein07:simpl_ion_heatin_rate_measur}.

Except for analytical results for some high symmetry configurations
\cite{reichle06:networ_surfac_elect_ion_traps}, modeling of SE traps
has so far been based mostly on numerical techniques.
While this approach is feasible for modeling with the purpose
of characterizing the trap provided by a given electrode structure, it
is inconvenient for modeling with the purpose of designing traps to
best meet design criteria.
This is unlike the situation in traditional rf traps such as the
linear Paul trap \cite{paul90:electromagnetic}, where the electrodes
surround the ion, and where symmetry considerations together with
scaling relations often provide sufficient background for trap
design.

In this paper, we present two methods for the design of SE
traps. Firstly, we demonstrate that a relatively unknown technique
makes it possible to analytically calculate the field resulting from
an arbitrary configuration of surface electrodes in the ``gapless
plane'' approximation where the electrodes are assumed to gaplessly
cover an infinite plane.
Secondly, we show that the important case of multipole guides with
translationally symmetric electrode configurations (analogs to the
linear Paul trap), can be given a simple geometrical description through a
conformal map.
Together with recent results on the fundamental constraints on
intersections in rf trap networks \cite{wesenberg08:intersecting}, the
results presented here can serve as guidelines in the design and
modeling of SE ion trap networks.

The paper is structured as follows: 
Sec.~\ref{sec:introduction} introduces the fundamental task
of ion-trap modeling.
In Sec.~\ref{sec:field-arbitr-electr}, we argue that SE traps can be
modeled as an infinite plane gaplessly covered by electrodes and
describe an analytical method for calculating the field of arbitrarily
shaped surface electrodes in this approximation.
In Sec.~\ref{sec:ring-trap} we apply this method to the analysis of
the SE ring trap.
Sections \ref{sec:fields-transl-symm} and \ref{sec:mp-depth} are
devoted to parametrizing and analyzing translationally symmetric SE
configurations.

\section{RF ion traps for QIP}
\label{sec:introduction}
\label{sec:surface-electrodes}

In this section we introduce the basic physics of rf traps and the
role of electrostatic modeling in the characterization of such traps.

For the typical dimensions ($\mu$m) and rf frequencies (MHz) of the rf
traps typically used for QIP, the rf field can be adequately treated
as quasistatic \cite{wineland98:exper_issues_coher_quant_state}.
Following Ref.~\cite{hucul07:transport} we introduce the electrode
basis function $\vbg{\gamma}(\vec{r})$ for an electrode $\gamma$ as
the unique function that takes the value $1$ ($0$) for $\vec{r}$ on
the surface of electrode $\gamma$ (all other electrodes) and fulfills
the Laplace condition $\nabla^2\vbg{\gamma}(\vec{r})=0$.
In terms of the electrode basis functions, the spatial potential
obtained by applying a voltage $\vag{\gamma}$ to electrode $\gamma$
while keeping all other electrodes grounded is
$\vsg{\gamma}(\vec{r})\equiv \vag{\gamma} \vbg{\gamma}(\vec{r})$.
Neglecting any nonlinear electric properties of the trap materials,
the total field $\vec{E}_{\text{Tot}}(\vec{r})$ in the trap can be
expressed as
\begin{multline*}
  \vec{E}_{\text{Tot}}(\vec{r},t) =
  -\left(\varf \cos(\Omega t)+\vacrf \right)\, \grad  \vbrf(\vec{r})\\  -
  \sum_i  \vaci{i} \grad \vbci{i}(\vec{r}), 
\end{multline*}
where $\Omega$ is the rf frequency, $\varf$ is the peak rf voltage,
$\vacrf$ is the bias applied to the rf electrodes, and $\vaci{i}$ is
the voltage applied to control electrode $i$. In most applications the
applied voltages will be modified quasistatically to manipulate the
trap, but we shall not be concerned with this in what follows.
To maintain physical dimensions, we will express most results in terms
of the spatial potential $\vsg{\gamma}(\vec{r})$ and the corresponding
field $\vec{E}_{\gamma}(\vec{r})\equiv -\grad \vsg{\gamma}(\vec{r})$.
Note that with this convention $\vec{E}_{\text{rf}}$ is the rf field
amplitude rather than the instantaneous rf field.

In an rf trap, the rf period $2\pi/\Omega$ is made to be the shortest
timescale of the system, so that in the adiabatic approximation the
effect of the rf field on an ion of charge $Q$ and mass $M$ is
described by the ponderomotive potential
\begin{equation}
  \label{eq:ppdef}
  \upp(\vec{r})\equiv\frac{Q^2}{4\,M\,\Omega^2} \left\lvert
    \vec{E}_{\text{rf}} (\vec{r}) \right \rvert^2, 
\end{equation}
corresponding to the kinetic energy of the micromotion induced by the
rf field \cite{dehmelt67:dehmel,ghosh95:ion}.
Including the effects of the quasistatic control fields, the effective
external potential experienced by a trapped ion is
\begin{equation}
  \label{eq:9}
  \ueff(\vec{r})\equiv
  \upp(\vec{r})
  + Q\left[\vscrf(\vec{r})+ \sum_i \vsci{i}(\vec{r})\right]. 
\end{equation}
Although trapping at the local minima of $\ueff$ where $\upp$ is
non-zero is possible, it is usually avoided as the micromotion
associated with a non-zero rf field amplitude can have a range of
detrimental effects.
In the following, we will only consider trapping at points where
$\vec{E}_{\text{rf}}(\vec{r})=\vec{0}$.

\section{Field of arbitrary surface electrodes}
\label{sec:field-arbitr-electr}

In the following, we argue that it is in many cases reasonable to
model an SE trap as an infinite plane gaplessly covered by electrodes.
Further, we show that in this approximation the field of an arbitrary
electrode can be calculated analytically.
Lastly, we consider the field in terms of the Fourier components of the
electrode shapes.

\subsection{Gapless plane approximation}
\label{sec:modeling-se-traps}

As introduced above, an SE trap has all electrodes located in a single
plane, typically on the surface of an insulating substrate as
illustrated in Fig.~\ref{fig:signe_elec}
\cite{chiaverini05:surfac_elect_archit_for_ion,seidelin06:microf_surfac_elect_trap_scalab,labaziewicz08:suppression}.

We will model the SE configuration in the gapless plane
approximation, where the gaps between individual electrodes are
assumed to be infinitely small and the electrodes are assumed to fully
cover an infinite plane with no other conductors above the plane.
It is clear that with $d$ denoting the ion-electrode distance,
necessary conditions for the gapless plane approximation to be valid
include that gaps between electrodes are much smaller than $d$, that
the extend of the trap is much larger than $d$, and that the distance
from ion to other conducting surfaces is much larger than $d$.

These requirements are often met in SE traps.
Firstly, although small electrode gaps lead to high field gradients,
SE traps are usually constructed with small interelectrode gaps since
any exposed surface of the substrate would be prone to pick up static
charges disturbing the operation of the trap.
The condition on the absence of other conductors in the trap region
imposes constraints on the experimental apparatus, but will in many
cases be met.
The corrections due to nearby conductors will be first order in the
ratio of the total electrode extend to the distance to the
conductors. For larger trap arrays, this could be a significant effect
and corrections should ideally be made.

% Numerical calculations indicate that in addition the aforementioned
% necessary conditions for the gapless plane approximation to be valid,
% we must require the total extent of the trap electrodes to be much
% smaller than the distance to other conductors. This requirement will
% often not be fully fulfilled, and corrections for the resulting
% finite-size effects should ideally be made.

In the gapless plane approximation, the electrostatic modeling for a
patch electrode $\gamma$, covering a region $A_\gamma$ of the
electrode plane reduces to identifying the unique potential
$\vsg{\gamma}(\vec{r})$ so that $\nabla^2\vsg{\gamma}(\vec{r})=0$ and
\begin{equation}
  \label{eq:seboundary}
  \vsg{\gamma}(x,y,0)=
  \begin{cases}
    \vag{\gamma}&\text{for $(x,y)\in A_\gamma$}\\
    0&\text{otherwise}, 
  \end{cases}
\end{equation}
where we have chosen the $x$-$y$ plane to coincide with the electrode
plane.

\subsection{Biot-Savart law}
\label{sec:biot-savart-law}

As shown in Ref.~\cite{oliveira01:biot_savar_like_elect}, the
electrical potential $\vsg{\gamma}(\vec{r})$ satisfying the boundary
condition of Eq.~(\ref{eq:seboundary}) can be very elegantly expressed
in terms of the solid angle $\Omega_{A_\gamma}(\vec{r})$ spanned by
$A_{\gamma}$ as seen from $\vec{r}$,
\begin{equation}
  \label{eq:biotsavartpot}
  \vsg{\gamma}(\vec{r})=\vag{\gamma}\, \frac{\Omega_{A_\gamma}(\vec{r})}{2\pi}.
\end{equation}
Taking the electrode plane to be the $x$-$y$ plane, and assuming $z>0$,
we have that
\begin{equation*}
  \Omega_A(\vec{r}) = \int_A \frac{
    \left(\vec{r}-\vec{r}' \right)\cdot \vhat{z}
  }{
    \left\lvert \vec{r}-\vec{r}' \right\rvert^3
  } dx'\,dy'.
\end{equation*}
The corresponding field takes the 
form \cite{oliveira01:biot_savar_like_elect}
\begin{equation}
  \label{eq:biotsavartfield}
  \vec{E}_\gamma(\vec{r})=\frac{\vag{\gamma}}{2 \pi} 
  \ointctrclockwise_{\partial A_\gamma}
  \frac{\vec{dr}' \times \left(\vec{r}-\vec{r}'\right)}
  {\left|\vec{r}-\vec{r}'\right|^3},
\end{equation}
where the integral is counterclockwise (as seen from above) along the
edge $\partial A_\gamma$ of $A_\gamma$. Note that the integral is a
Biot-Savart integral, so that $\vec{E}_\gamma$ is proportional to the
magnetic field that would be observed if a current was run along a
wire following the edge of $A_\gamma$. 
This points to a close connection to work on ``atom chips'' using
microfabricated wires
\cite{folman00:controlling,hansel01:bose-einstein,folman02:micros_atom_optic},
but it should be noted that this system is fundamentally different
since magnetostatic traps can trap at nonzero field minima without
adverse effects.
Numerical integration of
Eq.~(\ref{eq:biotsavartfield}) can benefit from the results of
Ref.~\cite{hanson02:compac_expres_biot_savar_field}.

\subsection{Fourier description of SE fields}
\label{sec:four-decomp-potent}

For some applications, e.g.~trap arrays \cite{chiaverini07:laserless},
it is useful to describe the surface electrodes in terms of their
Fourier transforms \cite{schmied08:sefourier}. This approach also
provides insight into the $z$ dependence of SE fields in general.

By employing the Dirichlet Green's function also used in
Ref.~\cite{oliveira01:biot_savar_like_elect}, we find that in terms of
$\vec{k}\equiv k_x \vhat{x}+k_y \vhat{y}$ the Fourier transform
$\tilde{V}(\vec{k},z) \equiv \int e^{-i \vec{k}\cdot \vec{r}}
\vsg{}(\vec{r})\, dx\,dy$ of $\vsg{}(x,y,z)$ with respect to the $x$ and $y$
coordinates for a constant $z$ value is given in terms of the surface
potential $\bar{V}_{\text{surf}}(x,y)\equiv \vsg{}(x,y,0)$ by 
\begin{align}
  \label{eq:13}
  \tilde{V}(\vec{k},z)
  &=\int e^{-i \vec{k}\cdot \vec{r}} \frac{1}{2\pi}
  \frac{\left\lvert \vhat{z}\cdot\vec{r} \right\rvert}{\left\lvert\vec{r}-\vec{r}'
    \right\rvert^3} \bar{V}_{\text{surf}}(x',y')
   \, dx\, dy\, dx'\, dy'\notag\\ 
  &= e^{- k \lvert z \rvert} \tilde{V}_{\text{surf}}(\vec{k}),
\end{align}
where $\tilde{V}_{\text{surf}}(\vec{k})\equiv \int e^{-i
  \vec{k}\cdot\vec{r}} V_{\text{surf}}(x,y)\, dx\, dy$ is the Fourier
transform of $V_{\text{surf}}$.
Equation (\ref{eq:13}) implies that the Fourier components of $V$ at a
given $z$ value fall off exponentially with $z$ on the length scale of
the electrode dimensions.

For our purpose, $\vsg{\text{surf}}$ will be equal to $\vag{}$ on some
region $A$ and zero elsewhere. In this case, $\tilde{V}_{\text{surf}}$
can be calculated by noting that since $e^{-i \vec{k}\cdot\vec{r}}$ is
equal to the $z$ component of the curl of the vector field
$\frac{i}{k^2} \left(\vhat{z}\times e^{-i \vec{k}\cdot \vec{r}}
  \vec{k}\right)$ and the vector triple product
$\vec{a}\times\vec{b}\cdot\vec{c}$ is invariant under cyclic permutations,
the Fourier transform can be converted to the path integral
\begin{equation*}
  \tilde{V}_{\text{surf}}(\vec{k})=
  \vag{} \frac{i}{k^2} \,\vhat{z}\cdot 
  \ointctrclockwise_{\partial A} e^{-i \vec{k}\cdot\vec{r}} \vec{k}\times\vec{dr},
\end{equation*}
where the integral is counterclockwise along the edge ($\partial A$) of $A$.

\section{Ring trap}
\label{sec:ring-trap}

\begin{figure}
  \centering
  \includegraphics[width=0.8\linewidth]{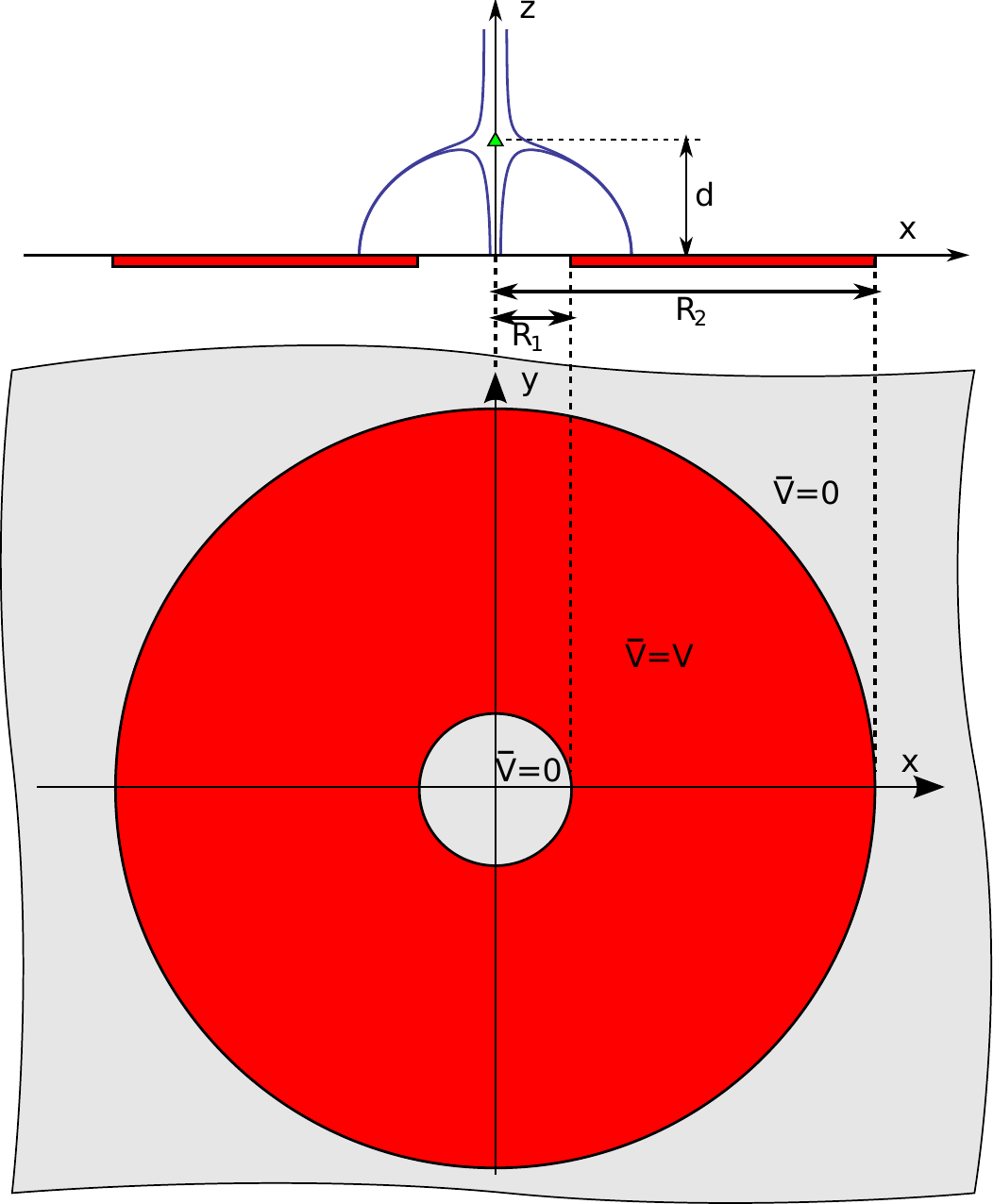}
  \caption{(Color online)
    Strongest possible SE ring-trap configuration.
    Upper part: Cut plane through the rotational symmetry ($z$)
    axis, together with the trap center (triangle) at a distance $d$ above
    the electrode plane, cut of ring electrode
    (rectangles below the $x$ axis), and field lines. 
    Lower part: Top view of the electrode ($x$-$y$) plane, which in the
    gapless plane model is assumed to be grounded everywhere but
    on the ring electrode with inner (outer) radius $R_1$ ($R_2$).  
  }
  \label{fig:ringtrap}
\end{figure}

To illustrate the application of the electrostatic methods presented
in the previous section to rf trap analysis, we will, as an example,
consider the SE version of a quadrupole ring trap
\cite{paul90:electromagnetic}. We will show below that this can be
implemented as a single ring-shaped rf electrode with inner and outer
radii $R_1$ and $R_2$, as illustrated by Fig.~\ref{fig:ringtrap}.

The field of the ring electrode at potential $\vag{}$ can be calculated as
the superposition of the fields of a disk of radius $R_2$ at potential
$\vag{}$ and a disk of radius $R_1$ at potential $-\vag{}$, so that the
combined field is  
$\vec{E}\equiv\vec{E}_{R_2}-\vec{E}_{R_1}$, where $\vec{E}_R$ is the field
of a disk of radius $R$ at potential $\vag{}$.
By symmetry, the electric field on the $z$ axis must be aligned along the
axis, and by Eq.~(\ref{eq:biotsavartpot}) we find that
\begin{equation}
  \label{eq:2}
  \vec{E}_R(z\, \vhat{z})
  = \vag{} \frac{R^2}{\left(R^2+z^2\right)^{3/2}}\, \vhat{z}.
\end{equation}
See, e.g., \cite{tryka99:method_for_calcul_averag_solid} for the off-axis
case.

For certain values of $R_1$ and $R_2$, we find that $E_z(z \vhat{z})$
vanishes for some value of $z$. 
When rf is applied to the ring electrode, such zeros correspond to
points of zero rf field amplitude, which provide an ideal trapping
point as discussed in Sec.~\ref{sec:introduction}.

Rather than trying to determine the position of any field zeros for
given dimensions of the ring electrode, we will consider the
constructive design problem of parametrizing all values of $R_1$ and
$R_2$ so that a field zero is located at a chosen distance $d$ above
the electrode plane.
As an intermediate step to achieving this, we parametrize $R_i$ as $d
\tan(\phi_i)$ so that
\begin{multline*}
  %\label{eq:ringtrapezofphiat1ms}
  E_z(d \vhat{z})=\frac{\vag{}}{d} \Big (
    \cos(\phi_1)\left[1-\cos^2(\phi_1)\right]\\
    -\cos(\phi_2)\left[1-\cos^2(\phi_2)\right]
  \Big ).
\end{multline*}
For $E_z(d \vhat{z})$ to vanish, $\cos(\phi_1)$ and $\cos(\phi_2)$
must be roots of the polynomial $u(1-u^2)-a$ for some
common constant $a$. This polynomial equation is in the Vieta standard
form \cite{birkhoff97:survey_mod_alg}, and the solutions of interest
to us can be parametrized as $\cos(\phi_{1,2})=2 \sin\left(\pi/6
\begin{smallmatrix}+\\-\end{smallmatrix} \theta\right)/\sqrt{3}$,
corresponding to $R$ values of
\begin{equation}
  \label{eq:ringtraprofphi}
  R_{1,2}=d \sqrt{
    \frac{3}{4} \sin^{-2}\left(\frac{\pi}{6}
      \begin{smallmatrix}+\\-\end{smallmatrix}
      \theta\right)-1
  }\quad \text{, $0<\theta<\tfrac{\pi}{6}$}.
\end{equation}
This, in particular, implies that the smallest outer radius of the ring
electrode is $R_2=d \sqrt{2}$.

\subsection{Ring -trap characteristics}
\label{sec:ring-trap-char}

\begin{figure}
  \centering
  \includegraphics[width=\linewidth]{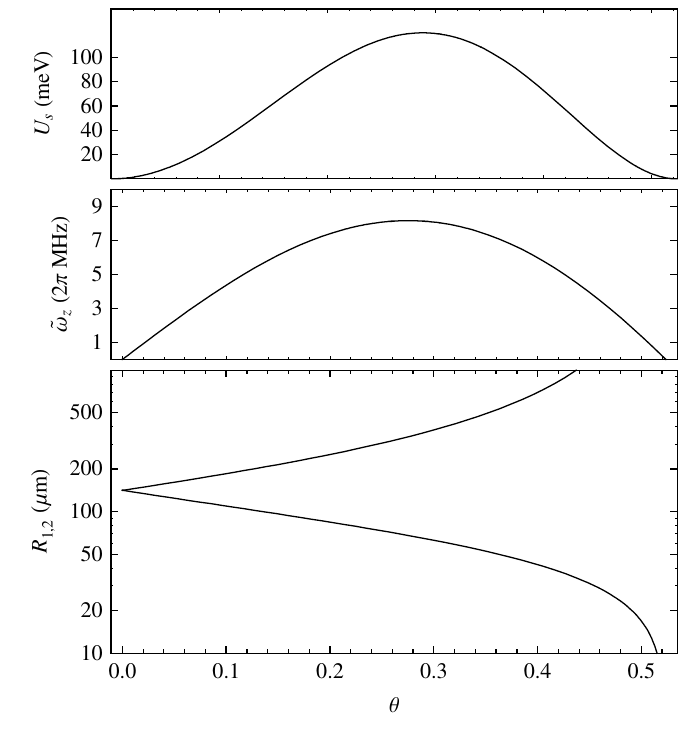}
  \caption{
    Properties of the SE ring trap as a function of the parameter
    $\theta$ appearing in Eq.~(\ref{eq:ringtraprofphi}) for the
    example parameter values given in Eq.~(\ref{eq:numbers}).
    $U_s$ (top) is the value of $\ueff$ at the saddle position in
    the absence of any control fields. 
    $\tilde{\omega}_z \equiv q_z \Omega / \sqrt{8}$ (middle) is the adiabatic
    approximation of $\omega_z$, valid when $q_z\ll 1$.
    $R_{1,2}$ (bottom) are the inner and outer radii of the ring
    electrode.
    The strongest trap 
    is obtained for $\theta\approx0.275$, where
    $\tilde{\omega}_z\approx 2\pi\times 8.17\,\text{MHz}$, 
    $U_s\approx118\,\text{meV},$ and $R_{1,2}\approx \{68,338\} \mu \text{m}$.
    Scaling of $q_z$, $U_s$, and $R_{1,2}$ are given by
    Eqs.~(\ref{eq:qnaughtdef}), (\ref{eq:usaddledef}), and $d$,
    respectively.  
  }
  \label{fig:ringtrapproperties}
\end{figure}

The different values of the parameter $\theta$ appearing in
Eq.~(\ref{eq:ringtraprofphi}) lead to traps with different
properties. In the following we consider the strength and the depth of
the trap, as illustrated in Fig.~\ref{fig:ringtrapproperties}.

The strength of an rf quadrupole trap is usually described by the
dimensionless parameter $q_z$ defined as \cite{ghosh95:ion}
\begin{equation}
  \label{eq:qnaughtdef}
  q_z\equiv 
  q_0
  \left\lvert 
    \frac{1}{2} 
    \frac{\partial^2 \vsrf(\vec{r})}{\partial z^2}
  \right\rvert \left(\frac{\varf}{d^2}\right)^{-1}, 
  \quad
  q_0\equiv \frac{4 \varf Q}{M \Omega^2 d^2},
\end{equation}
where $q_0$ has the same scaling properties as $q_z$.
In the adiabatic limit where the ponderomotive approximation is valid,
the secular oscillation frequency $\omega_z$ in the $z$ direction is
seen from Eq.~(\ref{eq:9}) to be $q_z \Omega/\sqrt{8}$.

For the ring trap the rotational symmetry and the Laplace condition
imply that the lowest-order multipole contributing at the field zero
must be a quadrupole term of the form
$\vec{E}(\vec{r})=-\alpha_z\topp{2} \left(2 z\vhat{z} - x \vhat{x}-y
  \vhat{y}\right)+\bigo(\vec{r}^2)$.
The quadrupole strength $\alpha_z\topp{2}$ can be calculated
as $\alpha_z\topp{2}=-\tfrac{1}{2}\left(\partial E_z/\partial
  z\right)$.
Evaluating the derivative of $\vec{E}_{R_2}-\vec{E}_{R_1}$ according
to Eq.~(\ref{eq:2}), we can write the resulting strength $q_z=q_0
\lvert\alpha_z\topp{2}\rvert d^2/\vag{}$ explicitly in terms of the parameter
$\theta$.
\begin{equation}
  \label{eq:ringStrength}
  q_z= q_0 \frac{1}{3}\left[\sin(5 \theta)-\sin(\theta) \right],
\end{equation}
which reaches a maximum value of $q_z\approx 0.473\, q_0$ when
$\cos^2(\theta)=(25+\sqrt{145})/40$.
For typical parameters of 
\begin{subequations}
  \label{eq:numbers}
  \begin{align}
    \Omega&=2\pi\times\, 100\, \text{MHz},&
    \varf &= 100\,\text{V},\\
    M&=10\,\text{AMU},&
    d&=100\,\mu\text{m},
  \end{align}
\end{subequations}
we find that $q_0 = 0.98$.

In addition to the trap strength, it is of practical interest to know
the depth of the trap, that is, the minimal energy required for a
trapped ion to escape.
To establish the depth we assume that the amplitude of the micromotion
induced by the rf field is small compared with the dimensions of the
potential. In this case, the depth of the trap region corresponds to
the height of the (lowest) saddle point, $\vec{r}_s$, of the trapping
(local) minima of $\ueff$.
We will for now ignore the control-field contributions to $\ueff$, and
only consider the ``intrinsic depth,'' $U_s\equiv \upp(\vec{r}_s)$
provided by the ponderomotive potential of the rf field.
By Eq.~(\ref{eq:ppdef}), $U_s$ is equal to
\begin{equation}
  \label{eq:usaddledef}
  U_s\equiv \
  U_0 \left\lvert
    \frac{\vec{E}_{\text{rf}}(\vec{r}_s)}{\varf/d} 
%    \frac{\grad \vsrf(\vec{r})}{\vag{}/d^2} 
  \right\rvert^2
  ,\quad  U_0 \equiv \frac{Q^2\,\varf^2}{4 M\,\Omega^2 d^2},
\end{equation}
where $U_0$ has the scaling properties of the depth.
As we shall see by example in Sec.~\ref{sec:modulated-depth}, control
fields can significantly change the effective depth of SE traps.

For the example parameters of Eq.~(\ref{eq:numbers}), $U_0 =
6.1\,\text{eV}$, which is not a typical depth: as illustrated by
Fig.~\ref{fig:ringtrapproperties} the maximal obtainable depth for a
ring trap corresponds to less than $2\%$ of $U_0$.

\section{SE multipoles}
\label{sec:fields-transl-symm}

Linear rf traps, such as the linear Paul trap, are based on an rf
multipole field, which is translationally invariant along the trap axis
thus providing a multipole guide. In addition to the ponderomotive
potential provided by the rf field, quasistatic control fields are
used to control the total trapping potential as illustrated in
Fig.~\ref{fig:signe_elec}.
With 3D electrode structures, the multipole guide can typically be
designed by symmetry considerations alone. In the linear Paul trap,
four parallel electrodes are used, an approach which can be extended
to higher-order multipoles \cite{ghosh95:ion}.
The same symmetry considerations are not applicable to
surface-electrode traps where only a few configurations of high
symmetry have been analyzed
\cite{chiaverini05:surfac_elect_archit_for_ion,reichle06:networ_surfac_elect_ion_traps}.

In the following, we aim to present a simple geometric theory of
multipole guides in SE traps, as well as analytical
methods for calculating the field in general translationally symmetric
electrode configurations.
The program will be similar to that followed in
Sec.~\ref{sec:ring-trap} for the ring trap: We will first introduce a
constructive parametrization of all possible SE multipoles with a
specific center position, and subsequently analyze trapping strength
and depth in terms of the design parameters. In this case we will also
briefly discuss the effects of biasing the rf electrodes.

\subsection{P{\'o}lya field}
\label{sec:polya-field}

\begin{figure}
  \centering
  \includegraphics[width=\linewidth]{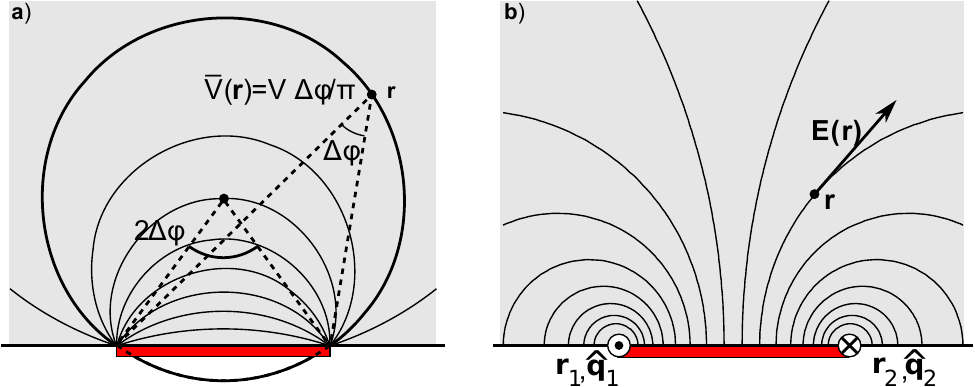} 
  \caption{(Color online)
    The potential (a) and field (b) above a strip electrode at
    potential $\vag{}$ (rectangle) with edges at $\vec{r}_1$ and
    $\vec{r}_2$ in an infinite grounded plane. 
    (a) 
    Since the solid angle spanned by the strip is $\Omega=2 \Delta
    \phi$, it follows from Eq.~(\ref{eq:biotsavartpot}) that the 
    equipotential lines are circular arcs through the
    strip edges as discussed in the text.
    (b) By Eq.~(\ref{eq:biotsavartfield}), the field is proportional
    to the magnetic field that would result from running a current
    along the edges of the strip in
    directions $\vhat{q}_{1,2}$. $\vec{E}(\vec{r})$ is given
    algebraically by Eq.~(\ref{eq:transfieldvector}).
    The field (potential) can be compactly expressed as a complex
    P{\'o}lya field (complex-valued potential) according to
    Eqs.~(\ref{eq:transfieldpol}) and (\ref{eq:transpotpol}).
  }
  \label{fig:stripfield}
\end{figure}

In the remainder of the paper, we will orient the coordinate system
so that the translational symmetry is along the $z$ axis and the
electrode plane is parallel to the $y$ axis at $x=d$. In figures we
will orient the coordinate system so that the origin (trap center) is
above the electrode plane, i.e., with the $x$ axis pointing downward.

It follows from Eq.~(\ref{eq:biotsavartfield}) that the field of a
strip electrode with edges intersecting the $x$-$y$ plane in
$\vec{r}_1$ and $\vec{r}_2$ is given by
\begin{equation}
  \label{eq:transfieldvector}
  \vec{E}(\vec{r}) = \frac{\vag{}}{\pi} \sum_{i=1}^2 \frac{
    \vhat{q}_i \times \left(\vec{r}-\vec{r}_i\right)
  }{
    \left\lvert \vec{r}-\vec{r}_i\right\rvert^2
  },
\end{equation}
for $\vec{r}$ in the trap region where $\vhat{q}$ is $\vhat{z}$
($-\vhat{z}$) for the left (right) edge as illustrated in
Fig.~\ref{fig:stripfield}(b).
In terms of the azimuthal angle $\Delta\phi$ spanned by the strip
electrode as seen from $\vec{r}$, the solid angle spanned by the
electrode is $\Omega=2 \Delta\phi$ so that by
Eq.~(\ref{eq:biotsavartpot}) $\vsg{}(\vec{r})=\vag{} \Delta \phi/\pi$.
This, in particular, implies that the equipotential lines are circles
through $\vec{r}_1$ and $\vec{r}_2$ as illustrated in
Fig.~\ref{fig:stripfield}(a). 

Although compact, the vector nature of Eq.~(\ref{eq:transfieldvector})
leads to somewhat unwieldy expressions for the combined field of
several strip electrodes.
A simpler representation can be obtained by mapping the point
$\vec{r}$ in the $x$-$y$ plane to the complex point $p$,
\begin{equation}
  \label{eq:pdef}
  p(\vec{r})\equiv x+i y,
\end{equation}
and the vector field $\vec{E}(\vec{r})$ to the complex function
\begin{equation}
  \label{eq:vecpolmap}
  \vecc{E}(p)\equiv E_x(\vec{r}(p))+i\,E_y(\vec{r}(p)).
\end{equation}
As discussed in more detail in Appendix \ref{sec:2d-electrostatics},
$\vec{E}$ is the P{\'o}lya field of $\vecc{E}^*$ and $\vecc{E}^*(p)$
is an analytical function of $p$ if and only if $\vec{E}$ is a
divergence-free and irrotational vector field.

In terms of $\vecc{E}$, the field of a strip electrode as given by
Eq.~(\ref{eq:transfieldvector}) is
\begin{equation}
  \label{eq:transfieldpol}
  \vecc{E}^*(z) =  \frac{\vag{}}{i \pi} \sum_{i=1}^2 \frac{q_i}{z-z_i},
\end{equation}
where $z=p(\vec{r})$, $z_i=p(\vec{r}_i)$, and $q_i\equiv \vhat{q}_i \cdot \vhat{z}$.
We have expressed $\vecc{E}^*$ in terms of $z$ rather than $p$ because
it turns out that Eq.~(\ref{eq:transfieldpol}) as well as
Eq.~(\ref{eq:transpotpol}) below hold without any modification for the
boundary conditions of a grounded cylinder ($z=c$) as we shall see in
the next section.

The complex potential $\Phi(z)$ fulfilling $\vecc{E}^*(z)=-\partial
\Phi(z)/\partial z$ is given by
\begin{equation}
  \label{eq:transpotpol}
  \Phi(z) =  \vag{} \frac{i}{\pi} \sum_{i=1}^2 q_i \ln(z-z_i),
\end{equation}
up to a complex constant. The real-valued physical potential $\vsg{}$
is equal to the real part of $\Phi$ as discussed in Appendix
\ref{sec:2d-electrostatics}.
For the purpose of calculating derivatives
Eq.~(\ref{eq:transpotpol}) is valid as it stands, but care should be
taken to choose the branch cut for the logarithm to ensure continuity
if the value of $\Phi$ is to be calculated directly.

\subsection{Parametrization of SE multipoles}
\label{sec:param-mult-guid}

% \begin{figure*}
%   \centering
%   \includegraphics[width=0.33\linewidth]{quadrupoleUnwrap_10}%
%   \includegraphics[width=0.33\linewidth]{quadrupoleUnwrap_17}%
%   \includegraphics[width=0.33\linewidth]{quadrupoleUnwrap_20}
%   \caption{Gradual unwrapping of a cylinder multipole.}
%   \label{fig:unwrap}
% \end{figure*}

\begin{figure}
  \centering
  \includegraphics[width=\linewidth]{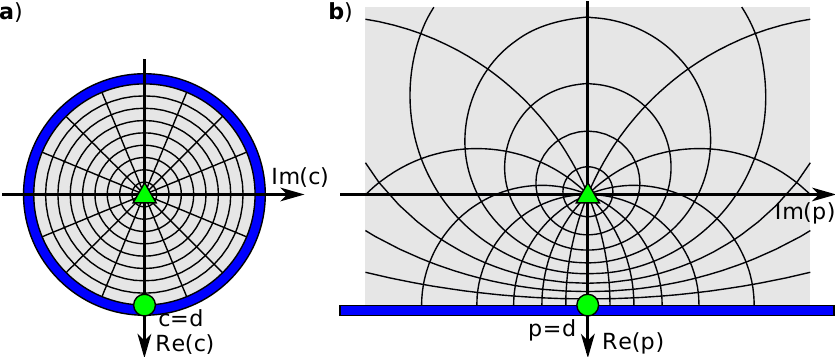}
  \caption{(Color online)
    A graphical representation of the conformal map $c\to p(c)$
    mapping the inside of a conducting cylinder to the trapping region
    above an infinite conducting plane, as defined formally by Eq.~(\ref{eq:stdmap}).
    Green circles  and triangles indicate the fixpoints at the point on
    the electrode plane right below the trap center ($c=p=d$) and at the trap
    center ($c=p=0$), respectively.
    Thick blue lines indicate the electrode surface in the cylinder
    picture (a) and the SE picture (b).  }
  \label{fig:map}
\end{figure}

\begin{figure}
  \centering
  \includegraphics[width=0.8\linewidth]{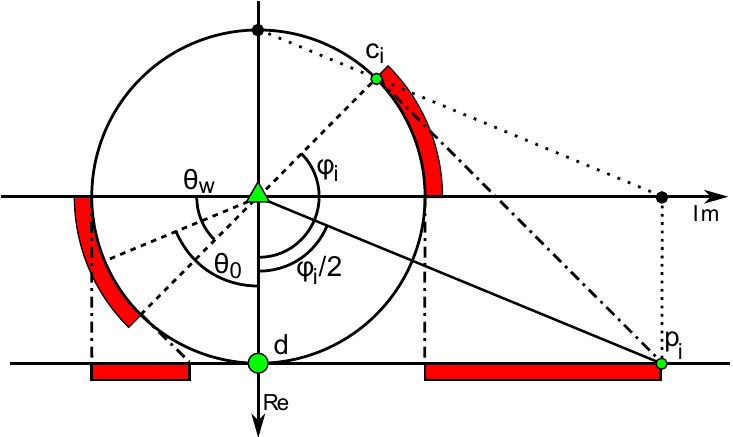}
  \caption{(Color online)
    Parametrization of multipole electrodes and geometrical
    implementation of the map Eq.~(\ref{eq:stdmap}) for the electrode
    surface.
    The plot shows the electrodes in the $c$ and $p$ representation
    superimposed. 
    Electrode positions for SE multipoles are specified by the
    parameters $n$, $\theta_0$, and
    $\theta_w$ according to Eq.~(\ref{eq:multipoleparamdef}). The
    example configuration 
    corresponds to that of Fig.~\ref{fig:quadexamples}(b).
    An arbitrary  point $c_i=d \exp(i \phi_i)$ on the electrode plane of the
    cylinder is mapped to $p_i=p(c_i)$ on the SE electrode plane
    $\re(p)=d$ according to Eqs.~(\ref{eq:stdmap}) or
    (\ref{eq:edgemap}).
    The mapping can be performed geometrically by extending the
    circle tangent at $c_i$ to the line $\re(p)=d$ (dash-dotted), or
    by a Mercator projection to the imaginary axis (dotted).  }
  \label{fig:multipoleparam}
\end{figure}

\begin{figure*}
%   \newlength{\myfulllength}
%   \setlength{\myfulllength}{\linewidth/3-10pt}
%   \hspace*{10pt}%
%   \parbox[b]{0pt}{%
%     \includegraphics[width=\myfulllength,trim=30pt 0pt 0pt 0pt,clip=false]{fulltrap_1}%
%   }\parbox[b]{\myfulllength}{\raggedright\textbf{a})}%
%   \parbox[b]{0pt}{%
%     \includegraphics[width=\myfulllength,trim=30pt 0pt 0pt 0pt,clip=true]{fulltrap_2}%
%   }\parbox[b]{\myfulllength}{\raggedright\textbf{ b})}%
%   \parbox[b]{0pt}{%
%     \includegraphics[width=\myfulllength,trim=30pt 0pt 0pt 0pt,clip=true]{fulltrap_3}
%   }\parbox[b]{\myfulllength}{\raggedright\textbf{ c})}%
  \includegraphics[width=\linewidth]{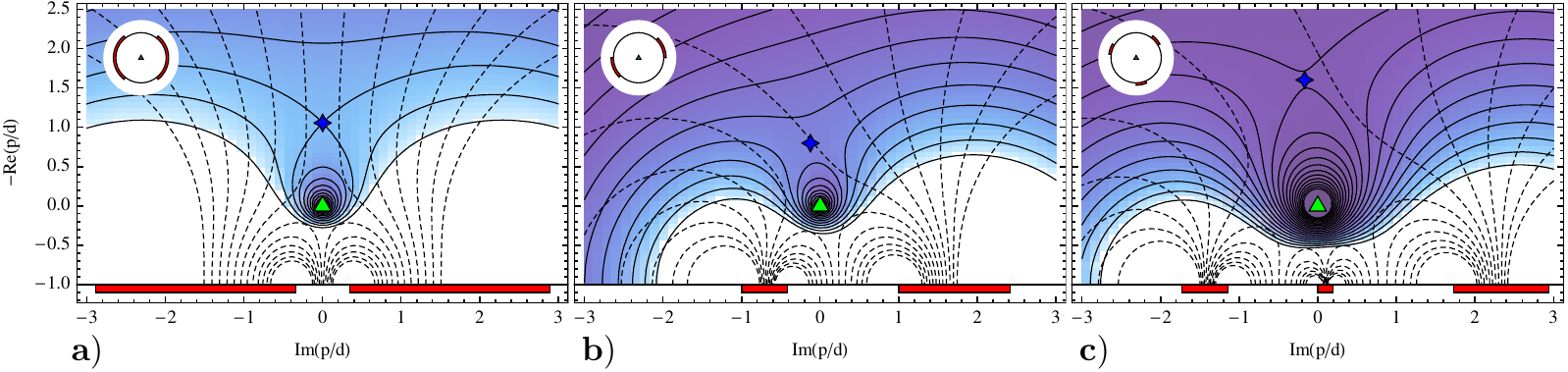}

  \caption{(Color online)
    Examples of SE multipoles. 
    Rectangles indicate rf electrode positions.
    Solid lines are contour lines of $\upp$, spaced by a
    factor of $\sqrt{2}$. The outermost line is at twice the maximal
    obtainable depth for quadrupole traps Eq.~(\ref{eq:quadusg}),
    corresponding to $2\times55.8\, \text{meV}$ for the example parameters 
    of Eq.~(\ref{eq:numbers}).
    Dashed lines are field lines (some omitted). Insets show the
    electrode configuration in the cylinder representation. Triangle
    and star marks indicate the guide center and intrinsic saddle,
    respectively.
    In terms of the parametrization Eq.~(\ref{eq:multipoleparamdef}), the
    configurations are described by 
    (a) the symmetric quadrupole ($n=2$) configuration ($\theta_0=\pi/2$) with
    $\theta_w$ chosen for optimal intrinsic depth according to
    Eq.~(\ref{eq:realsaddleconstraint});
    (b) a tilted quadrupole configuration with $\theta_w=\pi/4$
    and $\theta_0=5 \pi/8$; and
    (c) an octupole ($n=3$) configuration with $\theta_w=\pi/8$ and
    $\theta_0=\pi/16$.  }
  \label{fig:quadexamples}
\end{figure*}

With the P{\'o}lya field description of strip electrode fields provided by
Eqs.~(\ref{eq:transfieldpol}) and (\ref{eq:transpotpol}), identifying
the field zeros, i.e., trapping points, for a set of $n$ strip
electrodes is reduced to the problem of finding the roots of a
polynomial of order $2n-1$. Similarly, the task of characterizing
the identified trapping points in terms of multipole coefficients is
reduced to series expansion as discussed in Appendix
\ref{sec:2d-electrostatics}.

Although this can be achieved in many cases, we are often interested
in the inverse problem: how to place strip electrodes to achieve a
multipole guide of certain strength and orientation, and how these
parameters relate to total electrode area and other
implementation considerations.
To solve this, it is convenient to map the trapping region conformally
to a cylinder centered at the desired guide center, as an arbitrary
multipole in a cylinder with conducting walls can be designed by
symmetry considerations alone as detailed below.

We choose the trap center as the origin at a distance $d$ from the
electrode plane given by $x=d$, or equivalently, $\re(p)=d$, as
illustrated in Fig.~\ref{fig:map}.
A conformal map which maps the cylinder $\lvert c \rvert=d$ to the
electrode plane $\re(p)=d$ with fixpoints at c=p=0 (trap center) and
c=p=d (point on the electrode plane below the trap center) is given by the
conformal (M{\"o}bius) map
\begin{equation}
  \label{eq:stdmap}
  p(c)\equiv d \frac{2 c}{d+c},
\end{equation}
as illustrated in Fig.~\ref{fig:map}.

For the purpose of trap design, we are mostly concerned with the
mapping of the electrode structure.
Here, we find that a point $c=d \exp(i \phi)$ on the cylinder surface
is mapped to
\begin{equation}
  \label{eq:edgemap}
  \vec{r}(p(d\, e^{i \phi})) = d\, \vhat{x} + d\, \tan\left(\tfrac{\phi}{2}\right) \vhat{y},
\end{equation}
with a simple geometrical interpretation as illustrated in
Fig.~\ref{fig:multipoleparam}.

For the purpose of characterizing the SE multipoles, we will introduce
a parametrization of the simplest possible multipole guides where an
$n$th-order guide is implemented with $n$ strip electrodes placed
equidistantly around the cylinder walls.
We will parametrize this configuration by the angle of $\theta_w$
spanned by each strip as seen from the cylinder center, and the
azimuthal angle $\theta_0$ of the center of one strip,
so that the electrode edge positions are at $c=d \exp(i
\phi_n\topp{\pm})$ for $\phi_n\topp{\pm}$ given by
\begin{subequations}
  \label{eq:multipoleparamdef}
  \begin{align}
    \phi_n\topp{+}&=\theta_0 - \tfrac{\theta_w}{2} + 
    \tfrac{2\pi}{n} \left\{ 1, 2, \ldots, n \right\},\\
    \phi_n\topp{-}&=\theta_0  + \tfrac{\theta_w}{2} + 
    \tfrac{2\pi}{n} \left\{ 1, 2, \ldots, n \right\},
  \end{align}
\end{subequations}
as illustrated by Fig.~\ref{fig:multipoleparam}.

Together, Eqs.~(\ref{eq:edgemap}) and (\ref{eq:multipoleparamdef})
give the explicit position of the edges of strip electrodes to create
an $n$th-order SE multipole guide at a distance $d$ above the surface as
illustrated in Fig.~\ref{fig:quadexamples}.

\subsection{Field of SE multipoles}
\label{sec:prop-mult-guid}

To determine the analytical form of the field in an SE multipole
guide described by Eq.~(\ref{eq:multipoleparamdef}), we will 
calculate the field in the corresponding cylinder configuration
and then calculate the SE field through the map Eq.~(\ref{eq:stdmap}).

\begin{figure}
  \centering
  \includegraphics[width=0.5\linewidth]{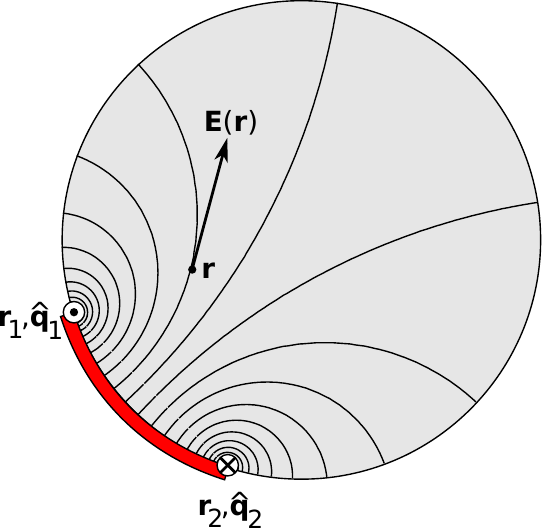}
  \caption{(Color online)
    The electric field of the field of a strip electrode at
    potential $\vag{}$ in a conducting cylinder has the exact same
    algebraic form as the field of a strip electrode in an infinite
    grounded plane, as given by Eq.~(\ref{eq:biotsavartfield}).
  }
  \label{fig:stripfieldcyl}
\end{figure}

Firstly, we will determine the field of a strip electrode with edges
at $c_1$, $c_2$ in a cylinder with conducting walls.
Calculating $\Phi(p(c))$ with $p_i=p(c_i)$ by
Eq.~(\ref{eq:transpotpol}) through the map Eq.~(\ref{eq:stdmap}), we
find that when considering pairs of edges with opposite sign $q_1+ q_2
=0$, $\Phi(c)$ is equivalent to the value obtained directly from
Eq.~(\ref{eq:transpotpol}) with $z=c$ and $z_i=c_i$, as illustrated by
Fig.~\ref{fig:stripfieldcyl}.
In other words, Eqs.~(\ref{eq:transfieldvector}),
(\ref{eq:transfieldpol}), and (\ref{eq:transpotpol}) for the field and
potential of strip electrodes in an infinite grounded plane hold
without any modification for the case of similar strip electrodes on
the walls of a grounded cylinder.
This is in agreement with the fact that 
the field of an SE strip electrode meets the boundary
conditions for a strip electrode in a conducting cylinder as
illustrated by Fig.~\ref{fig:stripfield}.

We can write the $\phi_n\topp{\pm}$ values of the electrode edges
Eq.~(\ref{eq:multipoleparamdef}) as $\phi_n\topp{0}+\theta_0\mp\theta_w/2$,
where $\phi_n\topp{0}\equiv 2 \pi \{1,2,\ldots,n \}/n$.  For
$\{c_i\}=d \exp(i \phi_n\topp{0})$ and $q_i=1$, we find that the sum
Eq.~(\ref{eq:transpotpol}) can be completed since $\prod_{k=1}^n \left( a-e^{i 2
    \pi k/n} \right)= a^n-1$, and the corresponding (unphysical)
single-sided complex potential is given by
\begin{equation*}
  \Phi_n\topp{0}(c)
%  \equiv  \vag{} \frac{i}{\pi} \sum_{n=1}^N  \ln\left(\tfrac{c}{d}-e^{i 2 \pi n/N}\right) 
  = \vag{} \frac{i}{\pi} \ln\left(\left(\tfrac{c}{d}\right)^N-1\right).
\end{equation*}
Since the positive and negative edge positions $\phi_n\topp{\pm}$ can
be obtained from $\phi_n\topp{0}$ by rotations through an angle of
$\theta_0 \mp \theta_w$ around the origin, the combined potential
$\Phi_n(c)$ is equal to $\Phi_n\topp{0}\left(c\,
  e^{-i(\theta_0-\theta_w/2)}\right) -\Phi_n\topp{0}\left(c\,
  e^{-i(\theta_0+\theta_w/2)}\right)$, which we write as
\begin{equation}
  \label{eq:multipolepot}
  \Phi_n(c)\equiv\vag{}\frac{i}{\pi} \ln\left(\frac{
      \left(e^{-i(\theta_0-\theta_w/2)} c\right)^n-d^n
      }{
      \left(e^{-i(\theta_0+\theta_w/2)} c\right)^n-d^n
      }
  \right).
\end{equation}
Equation (\ref{eq:multipolepot}) gives the exact form of a the complex
potential of a cylinder multipole as parametrized by
Eq.~(\ref{eq:multipoleparamdef}), and so together with the conformal
map Eq.~(\ref{eq:stdmap}) give the potential for an SE multipole in the
gapless plane approximation.
The physical potential with the specified boundary
conditions is $\re\{\Phi_n[c(p(\vec{r}))]\} + \vag{} n \theta_w/\pi$.

To learn about the strength and orientation of the corresponding
multipole field, we expand $\partial \Phi_n(c)/\partial c$ to lowest order about
the origin,
\begin{equation}
  \label{eq:multipoleexp}
%    \vecc{E}_n^*(c)
  \frac{\partial \Phi_n}{\partial c}
  =  
  e^{-i n \theta_0}
  \frac{\vag{}}{ d^n} 
  \frac{2 n}{\pi} 
  \sin\left(\frac{n \theta_w}{2}\right)
  c^{n-1}+\bigo(c^{2n-1}).
\end{equation}
Since by Eq.~(\ref{eq:stdmap}) $dc/dp=1/2$ at the origin, we have by
Eq.~(\ref{eq:multipolemap}) that the lowest nonzero multipole
coefficient of $\Phi_n$ is
\begin{equation}
  \label{eq:cylstrength}
  \alpha\topp{n}_p=2^{-n}\,\frac{2}{\pi}   
  \sin\left(\frac{n \theta_w}{2}\right) 
  \frac{\vag{}}{d^n}
  e^{-i n \theta_0},   
\end{equation}
with $\alpha\topp{n}_p$ as defined in Eq.~(\ref{eq:mpoleterms}).
The strength $\lvert\alpha_p\topp{n}\rvert$ is seen to reach its
maximal value of $2 \vag{}/ \pi (2d)^n$ for $\theta_w=\pi/n$,
corresponding to half the cylinder being covered with electrodes.
The complex phase factor $\exp(-i n \theta_0)$ describes the
orientation of the multipole.

At this point we can compare the strength of an SE multipole with that
of a traditional (3D) multipole guide with the same ion-electrode
distance.
The strongest such guide is obtained when the electrodes are chosen as
the $\pm \vag{}/2$ equipotential surfaces of $\re(z^n \vag{}/2 d^n)$, corresponding to
a multipole coefficient of
\begin{equation}
  \label{eq:strength3d}
  \alpha_{\text{3D}}\topp{n} = \frac{1}{2} \frac{\vag{}}{d^n},
\end{equation}
so that the strongest possible SE multipole is a factor of
$2^{n}\pi/4$ weaker than the strongest possible 3D multipole of the
same order and with the same ion-electrode distance.

For the case of quadrupole ($n=2$) guides employed in Paul traps, the
strength of the rf field is usually parametrized by the dimensionless
parameter $q$
\cite{paul90:electromagnetic,ghosh95:ion,wineland98:exper_issues_coher_quant_state},
which is related to the multipole expansion coefficients by
\begin{equation}
  \label{eq:qdef}
  q\equiv \left\lvert\alpha_p\topp{2}\right\rvert \frac{4\,
    \,Q}{M\,\Omega^2} = \frac{\sin(\theta_w)}{2 \pi} q_0, 
\end{equation}
for $q_0$ defined by Eq.~(\ref{eq:qnaughtdef}).  In the limit where
the adiabatic approximation is valid, the corresponding secular
frequency (in the absence of bias fields as discussed in
Sec.~\ref{sec:modulated-depth}), is well approximated by $q
\Omega/\sqrt{8}$ \cite{wineland98:exper_issues_coher_quant_state},
corresponding to a maximal secular frequency of $5.5\, \text{MHz}$ for
the example parameters of Eq.~(\ref{eq:numbers}).
To compare the results presented here with experimental results, we
note that Ref.~\cite{seidelin06:microf_surfac_elect_trap_scalab} gives
a measured value of $q_{\text{exp}}=0.54$ in the adiabatic
approximation for an SE trap with $\theta_w=\pi/4$, $\theta_0=\pi/4$,
$M=24\,\text{AMU}$, $\varf=103\,\text{V}$, and
$d=40\,\mu\text{m}$. This is in good agreement with the value $q=0.55$
predicted by Eq.~(\ref{eq:qdef}).

%\todo{compare numbers with Signes trap.}
%$\omega_{x,y}=15.7 MHz$
%$\Omega=2\pi\timex87\, \text{MHz}$,

\section{Depth of SE multipole traps}
\label{sec:mp-depth}

In this section we will investigate the potential depth that can be
obtained when using SE multipoles for rf traps.
As in Sec.~\ref{sec:ring-trap} on the SE ring trap, we will assume the
adiabatic approximation Eq.~(\ref{eq:9}) to be valid.
In Sec.~\ref{sec:mp-intrinsic-depth}, we will consider the
intrinsic depth, i.e., the depth in the absence of control fields. Here
we will mostly work with crude estimates, while Appendix
\ref{sec:saddl-mult-guid} provides a derivation of some exact results.
In Sec.~\ref{sec:modulated-depth} we consider the effects of control
fields on the total potential depth provided by $\ueff$.

Our current understanding of SE multipole depth is far from
complete and, in particular, Sec.~\ref{sec:mp-intrinsic-depth} contains
results based on conjectures as detailed in Appendix
\ref{sec:saddl-mult-guid}.

\subsection{Intrinsic depth}
\label{sec:mp-intrinsic-depth}

\begin{figure}
  \centering
  \includegraphics[width=\linewidth]{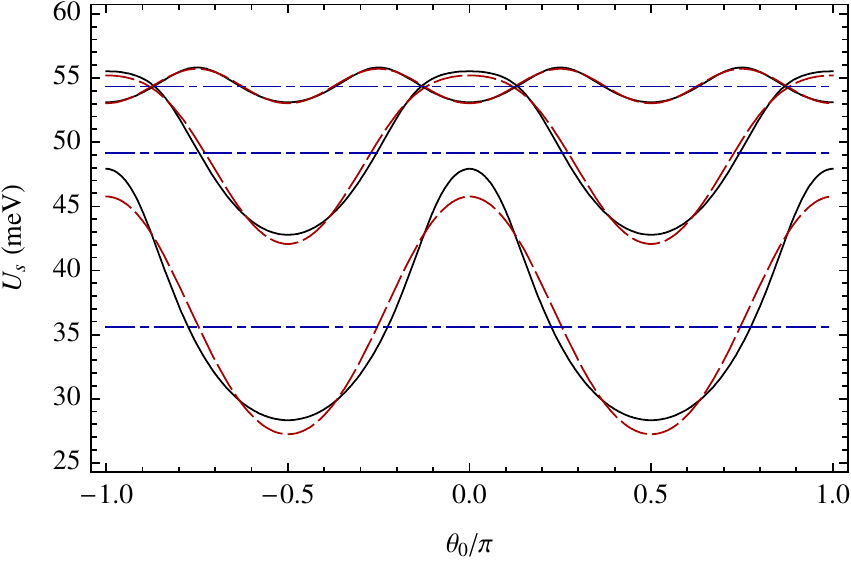}
  \caption{(Color online)
    Intrinsic depth of quadrupoles ($n=2$) as a function of $\theta_0$
    for $\theta_w/\pi$ equal to $0.3$, $0.4$, and
    $0.5$ with the parametrization described by
    Eq.~(\ref{eq:multipoleparamdef}) for the example parameters of
    Eq.~(\ref{eq:numbers}). The depth scales as $U_0$ (Eq.~(\ref{eq:usaddledef})).
    Solid lines are exact results, corresponding to cuts of the
    contour plot shown in Fig.~\ref{fig:saddleheight}.
    Dashed lines
    correspond to evaluating the exact field at the simplest saddle-point estimate
    $\pasaddle{2}{0}$ of Eq.~(\ref{eq:saddleapapprox}). Higher-order
    approximation as given by Eq.~(\ref{eq:piterate}), could not be
    discerned from the exact results.
    Dash-dotted lines show the value of the crude estimate of
    Eq.~(\ref{eq:cylusaddle}).  }
  \label{fig:saddleheightapprox}
\end{figure}

In terms of the complex rf potential $\Phi_{\text{rf}}$ the
ponderomotive potential is proportional to $\lvert \Phi_{\text{rf}}'
\rvert^2$ where we use a prime to denote differentiation with respect to
$p$. It follows that a necessary condition for an intrinsic saddle
point is that $\Phi_{\text{rf}}' \Phi_{\text{rf}}''=0$.
Since the zeros of $\Phi_{\text{rf}}'$ are not saddles, the saddle
point $\pbsaddle$ must fulfill $\Phi_{\text{rf}}''(\pbsaddle)=0$.

For the multipole configuration described by the complex potential
$\Phi_n$, we can take advantage of the fact that the expansion $\Phi_n
\approx \alpha_c\topp{n} c^n$ by Eq.~(\ref{eq:multipolepot}) is good
to order $2n-1$ to establish a reasonable estimate of the saddle point
$\psaddle{n}$ fulfilling $\Phi_n''(\psaddle{n})=0$.
By the chain rule
\begin{equation*}
  \label{eq:1}
  \Phi''\equiv
  \frac{\partial^2 \Phi}{\partial p^2}
  =\left(
  \frac{\partial^2 \Phi}{\partial c^2}
  -
  \frac{\partial \Phi}{\partial c}
  \frac{\partial^2 p}{\partial c^2}
  {\Big /} \frac{\partial p}{\partial c}
  \right)\,
  \left(\frac{\partial p}{\partial c}\right)^{-2},
\end{equation*}
so that for $\Phi\propto c^n$ we have a saddle at $c=d (1-n)/(1+n)$
corresponding to $p=d(1-n)$, which we take to be a reasonable estimate
of $\psaddle{n}$ as follows:
\begin{equation}
  \label{eq:saddleapapprox}
  \psaddle{n} \approx \pasaddle{n}{0} \equiv d(-n+1).
\end{equation}
The saddle point of $\upp$ for $\Phi_n$ is consequently located
approximately above the center of the guide at a distance of $n$ times
the ion-electrode distance from the electrode plane.
An iterative procedure for obtaining better estimates
$\pasaddle{n}{k}$, $k=1,2,\ldots$ of the true saddle $\psaddle{n}$ is
given by Eq.~(\ref{eq:piterate}).

As illustrated by Fig.~\ref{fig:saddleheightapprox},
$\upp\left(\pasaddle{n}{0}\right)$ is a reasonable estimate of the intrinsic
depth $\dsaddle{n}\equiv\upp\left(\psaddle{n}\right)$ corresponding to $\Phi_n$.
While $\upp$ can be evaluated exactly according to
Eq.~(\ref{eq:multipolepot}) by Eq.~(\ref{eq:phipofc}), the expression
is somewhat involved. 
A crude estimate of $\dsaddle{n}$ can be obtained by employing the 
approximation  $\Phi_n(p)\approx \alpha_c\topp{n}  c(p)^n$
again  to estimate the $\Phi'(\pasaddle{n}{0})$, resulting in 
\begin{equation}
  \label{eq:cylusaddle}
  \dsaddle{n} \equiv 
  U_0
%  \frac{q^2\, \vag{}^2}{4 M\,\Omega^2\, d^2} 
  \left\lvert
    \frac{\Phi_n'(\psaddle{n}) }{\vag{}/d}
  \right\rvert^2
  \approx
  \left[
  \frac{1}{n}
  \sin\left(\frac{n\, \theta_ w}{2}\right)
  \frac{4}{e^2 \pi}\right]^2 U_0,
\end{equation}
which for the example parameters of Eq.~(\ref{eq:numbers})
corresponds to $\dsaddle{n}\approx n^{-2} \sin^2\left(n\, \theta_ w/2\right)
180\, \text{meV}$.
The value of the crude estimate is illustrated in
Fig.~\ref{fig:saddleheightapprox}.

\begin{figure}
  \centering
  \includegraphics[width=\linewidth]{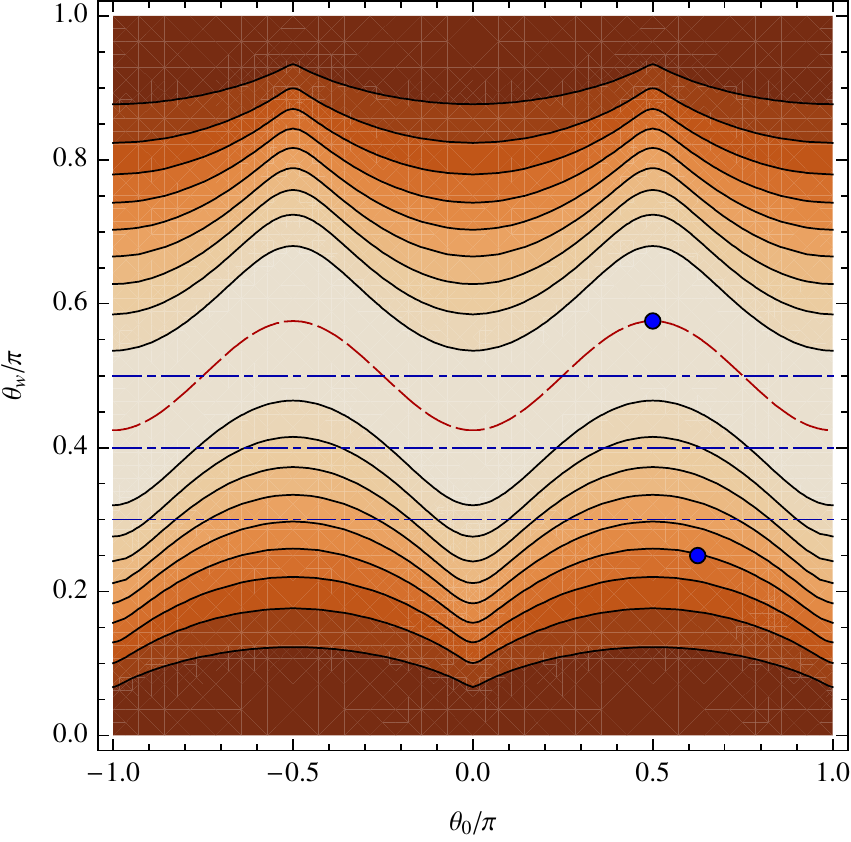}
  \caption{(Color online)
    Contour plot showing the intrinsic depth $\dsaddle{n}$ of quadrupole ($n=2$) 
    guides as a function of the parameters $\theta_0$ and $\theta_w$
    appearing in Eq.~(\ref{eq:multipoleparamdef}).
    Contour lines are spaced by $0.1$ times the maximal quadrupole
    depth $\dgsaddle{2}$ given by Eq.~(\ref{eq:quadusg}).
    The dashed line indicates points where
    Eq.~(\ref{eq:realsaddleconstraint}) holds. For these points the
    maximal depth $\dsaddle{n}=\dgsaddle{n}$ is obtained.
    The symmetries of the depth are discussed in Appendix
    \ref{sec:saddl-mult-guid}, and the dash-dotted lines (dots)
    indicate parameter values also illustrated in
    Fig.~\ref{fig:saddleheightapprox} [Figs.~\ref{fig:quadexamples}(a)
    and \ref{fig:quadexamples}(b)].
    }
  \label{fig:saddleheight}
\end{figure}

A more detailed analysis of the intrinsic depth is carried out in
Appendix \ref{sec:saddl-mult-guid}.
The main conclusion is that for any value of $n$ there is a special
saddle position $\pgsaddle{n}$ and a real-valued constant $A_n$, both
independent of the parameters $\theta_0$ and $\theta_w$, so
that $\psaddle{n}=\pgsaddle{n}$ if and only if $\theta_w$ and $\theta_0$ fulfill 
\begin{equation}
  \label{eq:realsaddleconstraint}
  \cos(n \theta_w/2)=A_n \cos(n \theta_0).
\end{equation}
Further, the intrinsic depth is the same for all combinations of
$\theta_0$ and $\theta_w$ fulfilling this requirement, and this depth
is the maximum obtainable in an $n$th-order SE multipole. To obtain
the maximal depth $\dgsaddle{n}\equiv\dsaddle{n}(\pgsaddle{n})$ and maximal strength simultaneously, we
must, by Eq.~(\ref{eq:cylstrength}), have $\theta_w=\pi/n$, so that
according to Eq.~(\ref{eq:realsaddleconstraint}), $\cos(n \theta_0)$
must also vanish. This corresponds to an antisymmetrical configuration
where a reflection in the real axis ($z$ axis) map rf electrodes to
ground electrodes and vice versa.

For quadrupoles ($n=2$),  $A_2=\sqrt{5}-2$ and the special saddle is located
at 
\begin{equation}
  \label{eq:quadpgsaddle}
  \pgsaddle{2}= -d\left(\sqrt{2+\sqrt{5}}-1\right).
\end{equation}
The maximal intrinsic depth of a quadrupole trap is 
\begin{equation}
  \label{eq:quadusg}
  \dgsaddle{2}\equiv \frac{5\sqrt{5}-11}{2\pi^2} U_0,
\end{equation}
with $U_0$ given by Eq.~(\ref{eq:usaddledef}). For the example
parameters of Eq.~(\ref{eq:numbers}) 
$\dgsaddle{2}=55.8\, \text{meV}$.
In the quadrupole case, the antisymmetric configuration
($\theta_0=\pi/2$, $\theta_w=\pi/2$) achieving maximal depth and
strength simultaneously is referred to as the four-wire trap
\cite{chiaverini05:surfac_elect_archit_for_ion,seidelin06:microf_surfac_elect_trap_scalab}.

\subsection{Depth in the presence of static fields}
\label{sec:modulated-depth}

The remainder of the section is devoted to a brief description of the
impact on static control fields on trap depth.  
For simplicity, we will not consider the effect of a confining
potential along the axis of translational symmetry.
Also, we only consider fields that vanish at the trap center, since
any field at this point would shift the trapping minimum of
$U_{\text{eff}}$ to a region of nonzero rf amplitude leading to
micromotion.

By applying different control voltages to different electrodes, a
control field of this type can be implemented with almost any set of
two or more electrodes.  Here, we will only consider the
possibility of applying a bias voltage $\vacrf$ to the rf
electrodes.
In practical applications, this approach has the advantage that a
control field generated by the rf electrodes is guaranteed to have a
field zero exactly at the trap center, even when departures from the
idealized gapless plane geometry are taken into account.
Also, this approach ensures that the principal axes of the control and
rf quadrupoles are aligned, allowing an analytical solution for the
motion of a trapped ion valid outside the adiabatic approximation
\cite{ghosh95:ion}.
The main practical obstacle to applying a bias to the rf field is that
it is incompatible with grounding part of the rf resonator.

For a bias voltage $\vacrf$ applied to the rf electrodes, the combined
effective potential is equal to
\begin{equation}
  \label{eq:ueffdcbias}
  \ueff(\vec{r})=
  U_0
  \left(
    \vdc \vbrf(\vec{r})
    +d^2\,\left\lvert 
      \grad \vbrf(\vec{r})
    \right\rvert^2
  \right),
\end{equation}
where $\vbrf(\vec{r})=\vsrf(\vec{r})/\varf$ is the rf electrode basis function and
\begin{equation}
  \label{eq:gdcdef}
  \vdc \equiv \frac{Q\, \vacrf}{U_0},
\end{equation}
parametrizes the strength of the bias field.
The ratio of the modified to the intrinsic depth only
depends on $\vdc$, so that the relative improvement of trap
depth, which can be achieved by biasing, is a purely geometrical factor
and does not depend on the operating parameters of the trap.

An important consideration when applying a control field is the effect
on trap stability. For the quadrupole trap with a control field
applied as a bias to the rf electrode, the stability criteria can be
established analytically \cite{paul90:electromagnetic,ghosh95:ion}.
For this purpose, the bias strength is parametrized by the
dimensionless parameter $a=2\, q\, \vacrf/\varf$ together with $q$
as defined by Eq.~(\ref{eq:qdef}).
The stable region usually used for QIP ion traps includes $(a,q)$ for
which $q<0.7$ and $\lvert a/q^2\rvert<0.5$
\cite{paul90:electromagnetic,ghosh95:ion}, where the second constraint
is the one of interest here.  If we introduce the dimensionless
``geometrical'' quadrupole strength
$\bar{\alpha}\topp{2}\equiv\alpha\topp{2} d^2/\vag{}$, so that
$q=\lvert\bar{\alpha}\topp{2}\rvert q_0$, we have that $a/q^2=\vdc/ 8
\lvert\bar{\alpha}\topp{2}\rvert$ so that the stability requirement is
a geometrical property independent of the operating parameters.
For the parametrization of Eq.~(\ref{eq:multipoleparamdef}) we have by
Eq.~(\ref{eq:cylstrength}) that
$\lvert\bar{\alpha}\topp{2}\rvert=\sin(\theta_w)/2\pi$ so that the
stability criterion is well approximated by $\lvert \vdc\rvert
\lesssim 0.6 \sin(\theta_w)$.

\begin{figure}
  \centering
%   \parbox{0.5\linewidth}{\raggedright\textbf{a})}%
%   \parbox{0.5\linewidth}{\raggedright\textbf{b})}\\
%   \includegraphics[width=\linewidth]{modulatedContours}
  \includegraphics[width=\linewidth]{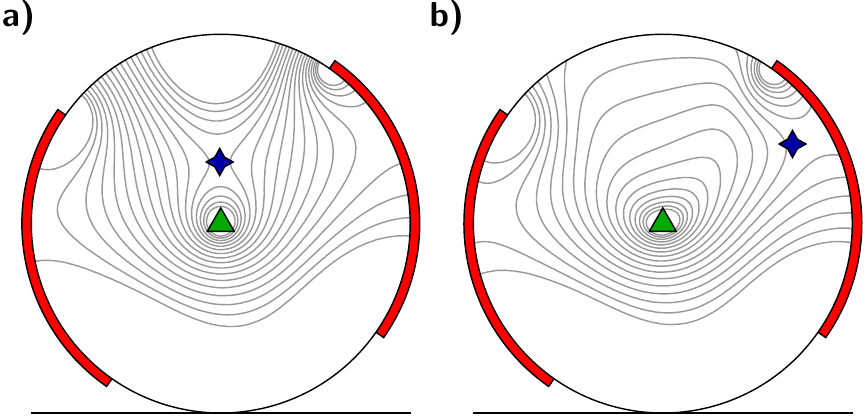}
  \caption{(Color online) Contour plot of $\ueff$ for the unbiased (a)
    and optimally biased (b) quadrupole configuration with 
    $\theta_0=100^\circ$ and $\theta_w=\pi/2$.
    In order to illustrate the potential in the full trapping region,
    the potential has been mapped to the $c$ plane so that what is
    plotted is $\ueff\left(p(c)\right)$ with $p(c)$ given by
    Eq.~(\ref{eq:stdmap}). Contours are spaced by a factor of
    $\sqrt{2}$ from $\dgsaddle{2}/4$ to 
    $64 \dgsaddle{2}$ with $\dgsaddle{2}$ as defined in
    Eq.~(\ref{eq:quadusg}).
    Stars indicate the lowest saddle points, and triangles indicate the trap centers.
    As discussed in the text, the relative increase in depth by a
    factor of $9.8$ is independent of operating parameters.  }
  \label{fig:modulatedcontours}
\end{figure}

As an example, we consider a quadrupole guide ($n=2$) with parameters
$\theta_0=100^\circ$, and $\theta_w=\pi/2$.
Here we find numerically that a maximum depth of $9.8\, \dgsaddle{2}$
is obtained for $\vdc=0.18$, corresponding to $a/q^2=0.14$, well
inside the stable region described above. The form of the optimally
biased $\ueff$ is illustrated by Fig.~\ref{fig:modulatedcontours}.
For the example parameters of Eq.~(\ref{eq:numbers}), the optimal
bias voltage is $\vacrf=\vdc U_0/Q = 1.1 \,\text{V}$.
Note that although the optimal bias voltage is dependent on operating
parameters, the ratio $a/q^2$ at optimal bias is constant, and so the
optimally biased configuration will always be stable, provided it
meets the constraint $q<0.7$ corresponding to a lower bound on
$\Omega$.

\section{Conclusion and outlook}
\label{sec:conclusion-outlook}

In conclusion, we have obtained a wide range of analytical results on
SE traps. The most significant are the explicit parametrization of SE
multipole geometry as given by Eq.~(\ref{eq:edgemap}), together with
the results on the exact field [Eq.~(\ref{eq:multipolepot})], strength
[Eq.~(\ref{eq:cylstrength})], and intrinsic depth [Eq.~(\ref{eq:quadusg}),
Figs.~\ref{fig:saddleheightapprox} and \ref{fig:saddleheight}].

There are, however,  a number of open questions relating to SE traps.

In direct continuation of the work presented here it would be of
interest to establish a formal proof that the intrinsic multipole
depth is maximal exactly when the saddle is located on the special
saddle point as conjectured in Appendix \ref{sec:saddl-mult-guid}.
Also, the adiabatic approximation will in many cases be invalid within
the stable region suggested by the intrinsic depth. It would
consequently be of interest to study the depth beyond the adiabatic
approximation \cite{mikosch07:evaporation}.

The limitations of the gapless plane approximation used
throughout this paper should be quantified in two respects.
Least importantly, the effect of finite gaps between electrodes are of
interest because increasing the gap size decreases field gradients and
the capacitive load of the trap.
More important are finite-size effects: In a realistic implementation
there will be grounded surfaces within a few cm of the electrode
surface.  Finite-size corrections will be first order in the total
electrode extend relative to this distance, and could consequently be
significant.

Finally, limitations on the possible global structure of SE fields are
not fully understood. In spite of recent progress on the possible
structure of intersections for rf trap networks
\cite{wesenberg08:intersecting}, it is not clear whether it is
possible to implement ideal intersections in SE traps.
Also, arrays of SE traps are being studied in several contexts
\cite{chiaverini07:laserless,schmied07:quantum}.

\begin{acknowledgments}
  The author gratefully acknowledges discussions with Jason Amini,
  Dietrich Leibfried, and David Wineland.
  This work was supported by the Danish National Research Agency, the Carlsberg
  Foundation, and the QIP IRC (Grant No. GR/S82176/01).
\end{acknowledgments}

\appendix

\section{Saddles in multipole guides}
\label{sec:saddl-mult-guid}

\begin{figure}
  \centering
  \includegraphics[width=\linewidth]{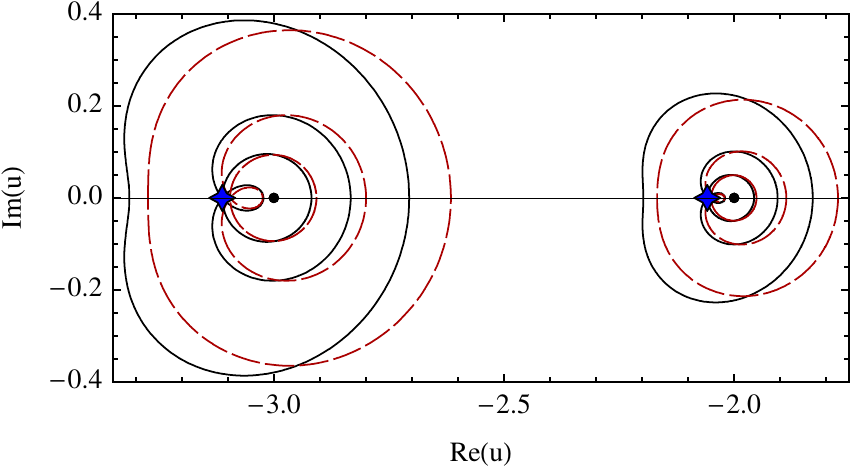}
  \caption{(Color online)
    Orbits of the intrinsic saddle position for $n=2$ and $3$ as a
    $\theta_0$ is varied for $n\theta_w/\pi=0.7$, $0.9$, and $1$. 
    Solid (dashed) lines show the exact value (the estimate $\usaddle{n,1}$) of 
    $\usaddle{n}$. Dots show the crude estimates $\usaddle{n,0}=-n$.
    Note that when $\theta_w$ is too large to be compatible with
    Eq.~(\ref{eq:andef}) for any value of $\theta_0$, the orbits do
    not pass through the special saddle positions $\ugsaddle{n}$
    (stars).  }
  \label{fig:saddleposition}
\end{figure}

This appendix is devoted to a detailed discussion of the exact
intrinsic depth and saddle position in SE multipole guides.
We use the term ``intrinsic saddle point'' to refer to a saddle point
of the ponderomotive part of $\ueff$ as given by Eq.~(\ref{eq:9}). As
discussed in Sec.~\ref{sec:mp-intrinsic-depth}, such saddle points
must be located at zeros of $\Phi_{\text{rf}}''$.
The following is devoted to identifying such zeros for
$\Phi_{\text{rf}}$ given by the exact multipole potential $\Phi_n$
of Eq.~(\ref{eq:ppdef}).
To make the algebra more manageable, we will set $\vag{}$ and $d$ to
unity, and introduce a new complex coordinate system with its origin in
the electrode plane.
\begin{equation*}
%  \label{eq:udef}
  u\equiv\frac{p}{d}-1.
\end{equation*}

With these conventions, $\Phi_n'=(\partial \Phi_n/\partial c)/
(\partial u/\partial c)$ can be given a compact form in terms of $c$
and  $\tilde{c}=e^{-i \theta_0}c$.
\begin{equation}
  \label{eq:phipofc}
  \Phi_n'=
  \frac{n}{\pi }\frac{\sin(n \theta_w/2)}{\tilde{c}^n-2\cos(n
    \theta_w/2)+\tilde{c}^{-n}}
  \frac{(1+c)^2}{c}.
\end{equation}
In terms of $u$, we have that
$c=(1+u)/(1-u)$ so that
\begin{equation*}
   \Phi_n'=
   \frac{4 n}{\pi}  
   \sin\left(\frac{n \theta_w}{2}\right) 
   \frac{\left(1-u^2\right)^{n-1}}{P(u)},
\end{equation*}
where the polynomial $P(u)\equiv P_+(u)+i P_-(u)$ is described as a
sum of even and odd parts, $P_\pm$, given by
\begin{subequations}
%  \label{eq:pdef}
  \begin{align}
    P_+(u)\equiv& 
    \cos(n \theta_0) \left[(1-u)^{2n}+(1+u)^{2n}\right]\notag\\
    &-2 \left(1-u^2\right)^n \cos\left(\frac{n \theta_w}{2}\right),\\
    P_-(u)\equiv& 
    \sin(n \theta_0) \left[(1-u)^{2n}-(1+u)^{2n}\right].
  \end{align}
\end{subequations}
We see that $\Phi_n'$ has zeros of order $n-1$ at $u=\pm 1$,
corresponding to the multipole center and its mirror image in the
electrode plane. Also, there are $2 n$ first-order poles at the roots
of $P$, which are located at the electrode edges on the imaginary
axis.

To establish the saddle positions, we differentiate $\Phi_n'$ again, and write
the derivative as
\begin{equation*}
  \label{eq:17}
  \Phi_n''= -\Phi_n' \frac{1}{P(u) (1-u^2)}
  S(u),
\end{equation*}
where 
$S(u)\equiv S_+(u)+i S_-(u)$ is a polynomial of order $2n+1$ with even
or odd parts given
by
\begin{equation*}
  \label{eq:7}
  S_{\pm}(u)=2 u (n-1) P_{\pm}(u)+(1-u^2) P'_{\pm}(u), 
\end{equation*}
so that, in particular,
\begin{multline*}
  S_{-}(u) =-2\sin(n \theta_0)\big( 
    u    \left[(u-1)^{2n} - (u+1)^{2n} \right]\\ 
    +n \left[(u-1)^{2n} + (u+1)^{2n} \right]
  \big).
\end{multline*}
We see that $\Phi_n''$ has zeros of order $n-2$ at $u=\pm 1$ and
second-order poles at the electrode edges, while the $2n+1$ roots of
$S$ describe the possible saddle positions.

To determine the root structure for $S$, we note the following:
According to the edge-current picture, the field must be antisymmetric
under inversion in the electrode plane so that
$\Phi'(u)^*=\Phi'(-u^*)$.
This implies that $\Phi'$ is real on the imaginary axis, or
equivalently, that the electrode field is perpendicular to the
electrodes at the surface as it should be.
Since $\Phi'$ will diverge with equal sign at the two sides of an
electrode or ground strip, there must be a zero of $\Phi''$ on each
such segment on the imaginary axis.
Further, since these zeros must also be roots of $S(u)$, there will be
a zero of $S$ between each of the $2n$ edges for a total of $2n-1$
zeros on the electrode plane, i.e., the imaginary axis.
As $S(u)$ is a polynomial of order $2n+1$ this leaves two roots
outside the electrode plane, which we identify as the saddle interest
to us, which we will denote $\usaddle{n}$ and for which
$\re(\usaddle{n})<0$, and its mirror image $-(\usaddle{n})^*$.

It follows from the above discussion that the saddle position
$\usaddle{n}$ corresponding to $\Phi_n$ as given by
Eq.~(\ref{eq:multipolepot}) is the unique root of $S(u)=0$ with
$\re(u)<0$. 
The roots of $S$ can be calculated numerically by standard
techniques, allowing a simple procedure for determining the intrinsic
saddle points for any multipole. 
Alternatively, the following heuristic iterative procedure
\begin{subequations}
  \begin{align}
    \label{eq:piterate}
    \uasaddle{n}{0}&=-n,\\
    \uasaddle{n}{k+1}&= n\, \uasaddle{n}{k} + \frac{1-u^2}{2} \frac{P'(u)}{P(u)},
  \end{align}
\end{subequations}
found by solving $S(u)=0$ for the first order $u$, has been observed
to converge relatively quickly to the main saddle as illustrated in
Fig.~\ref{fig:saddleposition}.
In particular, for estimating the trap depth, the zeroth- or
first-order estimates established by evaluating $\Phi'_n$ at
$\uasaddle{n}{0}$ or $\uasaddle{n}{1}$ are mostly sufficient, as
illustrated in Fig.~\ref{fig:saddleheightapprox}.
The initial estimate $\uasaddle{n}{0}$ is that identified in
Sec.~\ref{sec:mp-intrinsic-depth}, but we note that it could also be
obtained by starting the iteration at the trap center, $u=-1$.

\subsection{Saddle on the real axis}
\label{sec:saddle-real-axis}

\begin{table}[t]
  \centering
  \caption{\label{tab:sadles}%
    Numerical values of the special saddle point
    $\ugsaddle{n}$ and the factor $A_n$ appearing in
    Eq.~(\ref{eq:andef}), describing the relation between $\theta_0$ and
    $\theta_w$ for optimal depth.
    Exact values for $n=2$ are given by Eq.~(\ref{eq:quadpgsaddle}).
  }
  \begin{tabular*}{\linewidth}{@{\extracolsep{\fill}}lrr}
    \hline\hline
    $n$ & $-\ugsaddle{n}/n$ & $A_n$\\
    \hline
    % 2       &$\tfrac{1}{2}\sqrt{2+\sqrt{5}}$   & $\sqrt{5}-2$\\
    2	&1.02909	& 0.236068 \\           
    3	&1.03742	&-0.266149\\
    4	&1.04044	& 0.276076 \\
    10	&1.04375	& 0.286475 \\
    20	&1.04422	& 0.287935 \\
    50	&1.04436	& 0.288342 \\
    100	&1.04438	& 0.288400 \\
    200	&1.04438	& 0.288415 \\
    \hline\hline
  \end{tabular*}
\end{table}

It turns out that configurations where the saddle is on the real axis,
i.e., directly above the trap center, have a special significance.

There are two types of symmetry that will guarantee the saddle to be
on the real axis.
Firstly, when $\theta_0$ is equal to an integer multiple of $\pi/n$,
the electrode configuration is symmetric under reflection in the real
axis, ensuring that the saddle must be located on the axis.  In this
symmetric configuration, $S_-(u)$ is identically zero so that
$S=S_+$.
A less obvious symmetry is found when $\theta_w=\pi/n$ and
$\theta_0$ is an odd multiple of $\pi /2 n$, so that the electrode
structure is antisymmetric under reflection in the real axis, i.e., rf
electrodes are mapped to grounded areas and vice versa.
Since the ponderomotive potential only depends on the field strength,
this symmetry will also conserve the saddle which must consequently
be located on the real axis.

The antisymmetric configuration turns out to be of special interest,
and we will refer to the saddle position in this case as
$\ugsaddle{n}$.  $\ugsaddle{n}$ can be calculated as the negative real
root of $S_-$, since in the antisymmetric configuration $\cos(n
\theta_0)$ and $\cos(n \theta_w/2)$ both vanish so that $S=i S_-$.

Although we have introduced $\ugsaddle{n}$ as the saddle position for
the anti-symmetric configuration, so that in particular $\ugsaddle{n}$
is independent of $\theta_0$ and $\theta_w$, it turns out that there
is a number of configurations with the same saddle position, and that
these have very interesting properties.
Firstly, we note that since $P_+$ and thus $S_+$ depends on $\theta_0$
and $\theta_w$ only through the ratio of $\cos(n \theta_w/2)$ to
$\cos(n \theta_0)$, there will for any value of $n$ be a constant
$A_n$ so that $S_+(\ugsaddle{n})=0$ whenever
\begin{equation}
  \label{eq:andef}
  \cos(n \theta_w/2)=A_n \cos(n \theta_0).
\end{equation}
Since the roots of $S_-$ are independent of $\theta_0$ and $\theta_w$
it follows that when the condition of Eq.~(\ref{eq:andef}) is met, the
special saddle $\ugsaddle{n}$ will be a root both of $S_+$ and
$S_-$. Since $S=S_++i S_-$ it follows that $\ugsaddle{n}$ will also be
a root of $S$ and consequently be the actual saddle position through
an ``accidental'' symmetry as illustrated in
Fig.~\ref{fig:saddleposition}.
We have not been able to obtain closed expressions for $\ugsaddle{n}$
for $n>2$, but give numerical values in Table \ref{tab:sadles}.

The key property of the special saddle point is that all
configurations which fulfill Eq.~(\ref{eq:andef}), and thus have the
saddle at $\ugsaddle{n}$, obtain the maximal intrinsic depth possible
for a SE multipole of order $n$.
\begin{equation*}
%  \label{eq:conjecture}
  \dsaddle{n}(\theta_w,\theta_0)=\dgsaddle{n} \Leftrightarrow 
  \usaddle{n}(\theta_w,\theta_0)=\ugsaddle{n}.
\end{equation*}
This result has only been proven for $n=2$ but we conjecture it to be
true for all $n$ based on numerical results.

\section{2D electrostatics}
\label{sec:2d-electrostatics}

This section recapitulates a number of results on 2D electrostatics.

\subsection{P{\'o}lya field}
\label{sec:electr-compl-plane}

The P{\'o}lya field gives a one-to-one mapping from 2D electrostatic
fields to analytical functions on the complex plane \cite{polya:74} .
We consider the canonical mapping $(x,y) \mapsto z\equiv x+i y$ from
$\fieldR^2$ to $\fieldC$ and map the electrical field $\vec{E}(x,y)$
to the complex-valued function
\begin{equation}
  \vecc{E}(x+i y) \equiv E_x(x,y) + i E_y(x,y).
\end{equation}
Note that $\vecc{E}$ is not related to the complex amplitude used in
electrodynamics to describe phase relations.

We will consider $z\equiv x+i y$ and $\bar{z}\equiv
x-i y$ to be independent variables with respect to differentiation, so
that $\partial_z\equiv \tfrac{1}{2} \partial_x - i
\tfrac{1}{2} \partial_ y$ and
 $\partial_{\bar{z}} \equiv \tfrac{1}{2} \partial_x + i \tfrac{1}{2} \partial_y$ are the Wirtinger
derivatives often used in physics.
With this convention we find that
\begin{equation}
  \label{eq:laplacecond}
  \frac{\partial \vecc{E}}{\partial z}
  =\frac{\partial \vecc{E}^*}{\partial \bar{z}}
  =\frac{1}{2}\left( \grad\cdot\vec{E} + i\, \vhat{z}\cdot\grad\times\vec{E} \right),
\end{equation}
so that $\grad\cdot\vec{E}$ and $\grad\times\vec{E}$ vanish, as
they must for an electrostatic field in free space, if and only if
$\partial \vecc{E}/\partial z=0$.
In terms of the Wirtinger derivatives, a complex function $f(z)$ is
differentiable if and only if $\partial f(z)/\partial \bar{z}=0$, and
Eq.~(\ref{eq:laplacecond}) consequently implies that $\vec{E}$ is a
free space electrostatic field if and only if $\vecc{E}^*$ is complex
differentiable and thus analytic.
In complex analysis, the field $\vec{E}$ is known as the P{\'o}lya
field of $\vecc{E}^*$.

In terms of the real-valued potential $V$ the analytical function
$\vecc{E}^*$ is seen to be given by 
\begin{equation}
  \label{eq:3}
  \vecc{E}^* = -\left( 
    \frac{\partial V}{\partial x} - 
    i \frac{\partial V}{\partial y} 
  \right)=
  -2 \frac{\partial V}{\partial z}.
\end{equation}
Note that since $V$ is real it is not analytic and especially not an
antiderivative of $-\vecc{E}^*/2$.
On the other hand, the analytic function  $-\vecc{E}^*$ has an
analytic antiderivative $\Phi(z)$, so that
\begin{equation}
  \label{eq:5}
  \vecc{E}^* = - \frac{\partial \Phi}{\partial z} 
  \text{, and }
  \frac{\partial \Phi}{\partial \bar{z}} =0.
\end{equation}
The antiderivative is unique up to the addition of a complex
constant. By inspection, we see that $V$ defined as $V(x,y) =
\re(\Phi(x+i y))$ does fulfill Eq.~(\ref{eq:3}).
The imaginary part of $\Phi$ is known as the harmonic conjugate of
$V$, and we note that since $\grad \im(\Phi) \cdot \vec{E}=0$, lines
of constant $\im(\Phi)$ are field lines of $\vec{E}$.

\subsection{Conformal maps}

In two dimensions, conformal, or angle-preserving, maps will map a
solution of the Laplace equation to another solution of the Laplace
equation in the sense that if $\nabla^2 V(r)=0$ and $M(\vec{r})$ is a
conformal map, then we find that $\nabla^2 V(M(\vec{r}))=0$.
In the complex plane a map $w\mapsto z(w)$ is conformal if and only if
the function $z(w)$ is analytic. 
If $\vecc{E}(z)$ is a divergence-free and irrotational field, so that
$\vecc{E}^*(z)=-\partial \Phi(z)/\partial z$ for some complex-valued
analytic potential $\Phi(z)$ and $z=z(w)$ is an analytical function,
$\Phi_w(w)\equiv\Phi(z(w))$ is also analytic so that by
Eq.~(\ref{eq:laplacecond}) the P{\'o}lya field of
\begin{equation}
  \label{eq:fieldmapping}
  \vecc{E}^*_w(w) \equiv - \frac{\partial \Phi_w(w)}{\partial w} 
  = \vecc{E}^*(z(w)) \frac{\partial z(w)}{\partial w},
\end{equation}
is indeed divergence-free and irrotational by the results of the last
section.

\subsection{2D multipole expansion}
\label{sec:2d-mult-expans}

In two dimensions, a naive specification of the $n$th multipole
moment of a harmonic potential $V$ would involve the $n+1$ distinct partial
derivatives of order $n$. Considering the constraints imposed by the
Laplace condition would, however, leave only two free parameters.
Working in the complex notation, this result is obvious, as the
multipole expansion of $V$ is nothing more than the series expansion
of a corresponding analytical potential $\Phi$ so that
$V=\re(\Phi)$.
We will write the multipole expansion around the origin as
\begin{equation}
  \label{eq:mpoleterms}
  \Phi(z)=\sum_{n=0}^{\infty} \alpha\topp{n} z^n.
\end{equation}

We will now consider the mapping of the multipole strengths: If
$\Phi(z) = \alpha\topp{n} z^n+\bigo(z^{n+1})$ and
$\Phi_w(w)\equiv\Phi(z(w))$, where $z=z(w)$ is a conformal map so that
$z(0)=0$, we see that $\Phi_w(w)= \alpha\topp{n} (\partial z/\partial
w)^n w^n + \bigo(w^{n+1})$ so that the lowest-order multipole moment
of $\Phi_w$ is
\begin{equation}
  \label{eq:multipolemap}
  \alpha_w\topp{n}=\alpha\topp{n} \left(\frac{\partial z}{\partial w}\right)^n,
\end{equation}
where $\partial z/\partial w$ is evaluated at the origin.

%\bibliography{../../../bib/share,../../../bib/books}
\providecommand{\href}[2]{#2}\begingroup\raggedright\endgroup
\end{document}